\documentstyle[preprint,aps]{revtex}

\begin{document}
\title{
 SUPERCURRENT STATES IN 1D FINITE-SIZE  RINGS }

\vspace{0.5cm}
\author {
 Vladimir A.\ Kashurnikov and Alexei I.\ Podlivaev}

\address{
Moscow State Engineering Physics Institute,
115409 Moscow, Russia }

\author{
Nikolai V.\ Prokof'ev and Boris V.\ Svistunov }

\address{
Russian Research Center "Kurchatov Institute", 123182 Moscow, Russia}

\maketitle

\begin{abstract}
We consider topological supercurrent excitations (SC) in 1D
mesoscopic rings. Under certain conditions such excitations are well-defined
except for (i) a tunneling between resonating states with clockwise and
anti-clockwise currents, which may be characterized by the amplitude $\Delta$,
and (ii) a
decay of SC assisted by phonons of the substrate, both effects being
macroscopically small.
Our approach being based on the hydrodynamical action for the phase field
and its
generalization to the effective Hamiltonian explicitly takes into account
transitions between the states with different topological numbers and turns
out to be very effective for the calculation of $\Delta$ and
estimation of the decay width of SC, as well as for the unified description of
all known 1D superfluid-insulator transitions.

Most attention is paid to the calculation of the macroscopic scaling of
$\Delta$ (the main superfluid characteristic of a mesoscopic system) under
different conditions: a commensurate system, a system with single
impurity, and a disordered system. The results are in a very good
agreement with the exact-diagonalization spectra of the boson Hubbard models.

Apart from really 1D electron wires we discuss two other important
experimental systems: the 2D electron gas in the FQHE state and
quasi-1D superconducting rings. We suggest some experimental setups
for studying SC, e.g.,  via persistent current measurements,
resonant electro-magnetic absorption or echo signals, and relaxation
of the metastable current states.
\end{abstract}

\section{Introduction}
\label{sec:1}

The problem we address here is a general one for the condensed matter
physics and arises each time when we deal with the sample of nontrivial
topology. Namely, if we write down the field operator in terms of
density and phase $\Psi (x) = \mid \rho (x) \mid ^{1/2} e^{i\Phi (x)}$,
then in a sample with the topology of a ring we may define the topological
excitations characterized by the winding number
\begin{equation}
I={1 \over 2\pi } \: \oint_{\Gamma } \nabla  \Phi \:d\vec{l} \; ,
\label{I}
\end{equation}
where the integration contour $\Gamma $ is around the ring hole.
Clearly, these excitations exist in any space dimension and for $I\ne 0$
carry current due to nonzero gradient of the phase field.
The stability of topologically excited states does depend on the
dimensionality and properties of the bulk system.
In the normal state of the 3D and 2D  liquid the relaxation time of the
SC due to disorder or crystal potential is extremely
short (of order of the transport scattering rate).  However below the
superconducting or superfluid phase transition the relaxation mechanism
of $I$ is only through the vortices, which are macroscopic objects
in a sense that their energy is system size dependent (the pairs of
vortices have finite energy, but their only effect on $I$ is to provide
short-time fluctuations in $I$ leaving the long-time average of the
topological quantum number unmodified). In the ground state or at a very low
temperature the probability of vortex creation by quantum or thermal
fluctuations can be ignored, and the integral (\ref{I}) is a true
invariant of the system.
Note also, that at finite temperatures below the critical point there may
be introduced a regularized operator of phase in terms of which $I$ is also
an exact (with macroscopic accuracy) constant of motion \cite{Sv}.

The above argument does not work in 1D, because one can not
define vortex in one spatial dimension.
It means that the dynamics of $I$ is defined
by microscopic parameters of the problem, and the bare
transition amplitude between different quantum numbers is independent of
the system size. Calculating  the relaxation of the topological excitations
using bare parameters of
the Hamiltonian we conclude  in agreement with the
the conventional wisdom that disorder or periodic potential
does not allow conserving current states in 1D.
The crucial point for the study presented here is that the SC
in the interacting 1D system are always "dressed" by density
fluctuations. Any attempt to change $I$ would result in
driving the wavefunction of the whole ring into orthogonal state
(that is $ \langle I \mid I^{\prime} \ne I \rangle \to 0 $ with macroscopic
accuracy
- see also below). Now, the renormalized amplitude connecting different
numbers $I$ {\it is} size-dependent for the lowest excited states.
Depending on the interaction parameters between the particles in the
original model, one can obtain arbitrarily long relaxation times for
the topological excitations.

In this paper we discuss the stability and relaxation of SC in 1D
interacting systems. To be definite, most of the time we consider
Bose systems. As is well known \cite{Hald81a}, the statistics and
interaction effects are indistinguishable in 1D, so that fermions are
exactly described as hard-core bosons, and spin chains are converted into
the interacting bosons by Holstein-Primakoff transformation.
Our main interest is in the low-energy excitation spectrum of the ring
in the superfluid phase,
including the energy splittings between the resonating SC and their
hybridization with the sound modes.

Experimentally SC may be observed in the response of the ring to the vector
potential. Suppose that at the initial moment of time the system is in
the ground state with some value of $I=I_G$. In the general case $I_G$ is a
function of the magnetic flux through the ring and the parity of the
total number of particles in the system: for bosons and odd parity fermion
systems in zero flux $I_G=0$; fermion ring with the even number of
particles is equivalent to a bosonic system with the half-quantum
of the magnetic flux through the ring. In this case the
ground state is degenerate between $I= 1$ and $I=0$. By switching on
instantaneously an integer number of flux quanta, $\phi = m \phi_0 =m\: hc/e$,
we  excite nonadiabatically the supercurrent state with $I =I_G+ m$.
One may then study the time evolution of $I$ by measuring e.g. the
oscillations and decay of the magnetic moment generated by the current. In
a static experimental setup the relevant quantity is the
magnitude and sign of the persistent
current near the points of the ground state degeneracy.
In bosonic  and
odd parity fermionic systems these points are given by
$\phi \approx (m+1/2) \phi_0$. Near these points
the persistent current has the maximum possible amplitude and abruptly
changes its sign. The width of the crossover region is very small
$\mid \phi -(m+1/2)  \phi_0 \mid \ll \phi_0$ and directly related
to the frequency of the current oscillations in the dynamic experiment.
In the last section we will describe in some detail the most
promising experimental 1D system - edge states of the 2D electron gas in
the regime of fractional quantum Hall effect (FQHE).
Here the SC are related to
the electrical charge transfer between the edges in the Carbino disk geometry
thus making it possible to study the phenomenon by  monitoring
the charging effects between the edges.
Finally, we will demonstrate that the physics of SC relaxation in 1D systems
can be directly applied to a 3-dimensional
superconducting ring with the ring length being much longer than the
wire cross-section diameter.

The other way to excite the SC is by pumping the system with the
high-frequency electromagnetic radiation in resonance with the energy
of, say,  the first  current state.

The paper is organized as follows. In section \ref{sec:2} we formulate
the effective-action approach to the problem and
show it to be very convenient for unified description of
the critical points and renormalization group (RG) equations
 for the superfluid-insulator transitions in commensurate and disordered
systems. Next we calculate the energy splitting, $\Delta_{AB}$, between
degenerate current groundstates (in the presence of the half-quantum of
magnetic flux or its equivalent) using an instanton technique. This
splitting and its dependence on the ring length $L$
is the main mesoscopic characteristic of SC. It gives the frequency of
the coherent current oscillations in the superfluid phase, while its
divergence signals about the superfluid-insulator transition. All
calculations in section \ref{sec:2} are done with the logarithmic
accuracy.

In section \ref{sec:3} we formulate an alternative approach
based on the effective Hamiltonian  for the SC dynamics. This way we
calculate precisely the hybridization coupling between the topological
states and phonons, which allows to fix all the ratios
between the lowest energy levels. The other goal is the
possibility to account for the resonances between the SC and sound waves
and describe correctly level crossings. All the results obtained
analytically in sections  \ref{sec:2} and \ref{sec:3} are tested
in section \ref{sec:4} by intensive numeric calculations done on
finite-size Hubbard-like ring models.
We present the low-energy spectra for various
systems and discuss the procedure of their extrapolation on larger
system size by combining the numeric data with the analytic RG equations.
In section \ref{sec:5} we discuss possible experimental
setups for the study of SC.
We also considered in some detail two
important experimental systems, i.e., the
2D electrons in the FQHE state and a long, narrow
3D superconducting ring, as the most promising
systems where our  results  may be tested.

\section{Effective-action approach}
\label{sec:2}

Since the stability of the SC states is associated with the
conservation of the topological invariant $I$ (\ref{I}) the most natural
language for the description of its violation or even complete disappearance
under various conditions is that which can directly take into account
processes changing the value of $I$. In this section we introduce such a
language in terms of the effective action for the phase field in (1+1)
dimensions.

Consider Popov's hydrodynamical action \cite{Pop} with the density-field
fluctuations integrated out:
\begin{equation}
S[\Phi ] = \int dx \: d \tau \left\{ in_0 (x) \Phi_{\tau}' + {1 \over 2}
\Lambda_s (\Phi_x')^2 + {1 \over 2}\kappa (\Phi_{\tau}')^2  \right\} \; .
\label{S1}
\end{equation}
Here $\Phi(x,\tau )$ is the phase field ($x$ is the spatial coordinate and
$\tau$ is the imaginary time), $n_0(x)$ is the local number density with the
coordinate dependence coming from this or that short-range external potential,
$\Lambda_s$ is the superfluid stiffness, and $\kappa = (d\mu / dn)^{-1}$ is
the compressibility ($\mu$ is the chemical potential and $n$ is the
macroscopic density). As we shall see below, the dependence of $n_0$ on $x$
in (\ref{S1}) is of prime importance being responsible for a particular type
of possible violation of superfluidity. So $n_0(x)$ can not be replaced by
$n$.

Zero-point fluctuations with the change of $I$ are represented by the
topological defects (vortices) in the field $\Phi(x,\tau )$. Hence, the
question of the existence in a ring system of the metastable groundstate-like
excitations with nonzero expectation values of $I$ is the question of
statistics of the vortices. If in the long-range limit the vortices are
absent (bounded in short-range pairs) such states are possible. If separate
vortices are present at the arbitrarily large scales one can not introduce
the topological quantum number and the only possible equilibrium expectation
value of $I$ is zero. In this case the very description of the system in
terms of the effective action (\ref{S1}) is, generally speaking, no longer
valid because of divergent long-range renormalizations.

The self-consistent criteria for superfluidity under different conditions
(commensurate external potential, disorder, or a special case with the single
impurity) follow directly from the action (\ref{S1}). With rescaled
variable $\tau$ this action reads
\begin{equation}
S[\Phi ] = \int dx \: dy \left\{ in_0(x) \Phi_y' + {1 \over 2\pi K}
(\nabla \Phi)^2 \right\} \; .
\label{S2}
\end{equation}
Here $y=c\tau$ ($c=\sqrt{\Lambda_s / \kappa}$ is the sound velocity),
\begin{equation}
K^{-1} = \pi \sqrt{\Lambda_s \kappa}\; .
\label{K}
\end{equation}
The first term in (\ref{S2}) is of topological nature and "reacts" only on
topological defects. Integrating  over $y$ and introducing
\begin{equation}
\gamma (x) = -2\pi \int_0^x n_0(x') dx'
\label{gamma1}
\end{equation}
one may rewrite $S[\Phi ]$ as
\begin{equation}
S[\Phi ] = {1 \over 2\pi K} (\nabla \Phi)^2  + i\sum_j p_j \gamma (x_j)
\; ,
\label{S3}
\end{equation}
where $j$ enumerates the vortices, $p_j=\pm 1,\pm 2, \ldots$ and $x_j$ are
the charge and the $x$-coordinate of the vortex $j$, respectively. The
effective action
(\ref{S3}) has the form of $XY$-model (the first term) with the additional
feature that each vortex brings in a spatially- and charge-dependent phase
(the second term). This phase plays an extremely important role. To realize
this, consider a homogeneous system. In the absence of the second term in the
action (\ref{S3}) the system would be equivalent to the $XY$-model and thus
would demonstrate Kosterlitz-Thouless  transition at $K=1/2$ which means
destruction of superfluidity. However, it is clear that the homogeneous system
is always trivially "superfluid" because of the momentum conservation. Hence,
no phase transition actually should occur, and it is the second term in
Eq.(\ref{S3}) that corrects the result. Indeed, in the homogeneous case
$\gamma (x_j)$ is a linear function of $x_j$ and the integration over $x_j$
makes the net contribution of the vortex $j$ to the partition function to be
zero. A similar picture takes place in a periodic incommensurate
potential (and, in particular, in a lattice at an incommensurate filling). In
this case $\gamma (x)$ may be represented as
\begin{equation}
\gamma (x) \equiv \gamma (s,\xi) = -2\pi \nu s - 2\pi \int_0^{\xi} n_0(x') dx'
 \; ,
\label{gamma2}
\end{equation}
where $x=sl+\xi$, $l$ is the period of the potential,
$s$ is the corresponding number of periods,
$\xi \in [0,l]$ is the reduced coordinate,
$\nu = \int_0^l n_0(x) dx \equiv nl$ is the filling factor. The second term in
the r.h.s. of Eq.(\ref{gamma2}) leads only to a certain short-range
renormalization of the statistical weight of the vortex and, thus, does not
play an essential role, so we omit this term from now on. (Note also, that in
a lattice this term is absent from the very beginning.) The first term in the
r.h.s. of Eq.(\ref{gamma2}) is linear in $s$. This means that just as in the
homogeneous case the net contribution of vortices to the partition function
is zero, provided $\nu$ is not a rational number. Hence, no phase transition
occurs.

In the case of commensurate filling the discreteness of $s$ becomes important.
Now there appears a class of vortices whose net contribution is nonzero.
These are the vortices whose charges are the multiples of $p$, the
denominator of the filling factor. All these vortices enter the partition
function with one and the same spatially independent phase, so one deals with
the $XY$-model with the constraint that the charge of a vortex be a multiple
of $p$. Thus, the system should experience a Kosterlitz-Thouless-type phase
transition at $K=p^2/2$, which can be identified with the known superfluid -
Mott insulator transition in a commensurate system \cite{Col,Hald81a,GS}.
Note, that the
critical value of $K$ follows immediately from the effective action (\ref{S3})
by the free-energy-sign argument for a single vortex \cite{KT}. This argument
is exact and need not a resorting to the RG, provided $K$ is understood as the
long-range asymptotic quantity \cite{Sv}. Nevertheless, RG equations are of
interest for some applications (see below). They read (cf. \cite{GS})
\begin{equation}
\left\{ \begin{array}{l}
dK / d\lambda = w^2 \\
dw / d\lambda = \left( 2 - p^2/K \right) w
\end{array} \right.
\label{RG1}
\end{equation}
Here $\lambda = \ln R$, $R=\sqrt{x^2+y^2}$ being the characteristic scale of
distance, $K(\lambda)$ is the mesoscopic value of $K$. Eqs.(\ref{RG1})
are the standard Kosterlitz-Thouless RG equations
\cite{KT,K} written for the vortices of the charge $p$. As usual,
$(w(\lambda))^2$ is
proportional to the number of vortex pairs of the size $\sim R$ in the area
$\sim R^2$ and is much less than unity in the region of applicability of
Eqs.(\ref{RG1}). Mesoscopic values of $\Lambda_s$ and $\kappa$
are related to$K(\lambda)$ by
\begin{equation}
\Lambda_s (\lambda) = \frac{c}{\pi K(\lambda)} \; , \;\;\;
\kappa (\lambda) = \frac{1}{\pi c K(\lambda)} \; ,
\label{rel1}
\end{equation}
$c$ being irrenormalizable quantity.

Considering disordered superfluid system it is convenient to rewrite the
second term in (\ref{S3}) in terms of the pairs of vortices:
$\sum_j p_j \gamma(x_j) \rightarrow \sum_{\zeta} q_{\zeta} \,
\tilde{\gamma}(x_{1\zeta},x_{2\zeta})$, where $\zeta$ indexes the vortex
pairs, $q_{\zeta}=1,2,3,\ldots$ is the charge of the vortices in the pair
$\zeta$, $x_{1\zeta}$ and $x_{2\zeta}$ are the $x$-coordinates of the positive
and the negative vortex in the pair $\zeta$, respectively, and
\begin{equation}
\tilde{\gamma}(x_1,x_2) = -2\pi \int_{x_1}^{x_2} n_0(x) dx \; .
\label{gam_til}
\end{equation}
Being interested in the long-range contribution of a vortex pair we may
average its phase factor over short-range translations along the $x$-axis:
\begin{equation}
e^{i2\pi q\, \tilde{\gamma}(x_1,x_2)} \rightarrow \langle
e^{i2\pi q\: \tilde{\gamma}(x_1+\xi , x_2+\xi )} \rangle _{\xi} = f(x_1-x_2)
\; .
\label{aver}
\end{equation}
Since $n_0(x)$ in a disordered system is a random function with some
microscopic correlation length, the function $f(x_1-x_2)$ decays at
microscopically small $\mid x_1-x_2 \mid$ and may be replaced by
$\propto \delta (x_1-x_2)$. We see thus, that in a disordered system the
second term in the effective action (\ref{S3}) is equivalent to the constraint
that only "vertical" vortex pairs be present. This constraint changes the
entropy per vortex as compared to the regular commensurate case. Now the
entropy per vortex is $(3/2) \ln L + \mbox{\it finite term}$ since the
entropy of
a pair is obviously $3 \ln L + \mbox{\it finite term}$. The energy of a vortex
with the minimal charge is $K^{-1} \ln L + \mbox{\it finite term}$, so the
free-energy-sign argument yields the critical macroscopic value of $K=2/3$,
in agreement with the result of Giamarchi and Schulz \cite{GS} obtained from
other considerations. The renormalization group for the disordered case
may be easily derived from the effective action in analogy with the
Kosterlitz-Thouless treatment \cite{KT}. The new feature associated with the
above-mentioned constraint is that renormalizable  quantities now are
$\Lambda_s$ and $c$ while $\kappa$ can not be renormalized by the vertical
vortex pairs. The RG equations read now
\begin{equation}
\left\{ \begin{array}{l}
dK / d\lambda = K w^2 \\
dw / d\lambda = \left(3/2 - 1/K \right) w
\end{array} \right.
\label{RG2}
\end{equation}
Here $\lambda$ and $w$ have the same meaning as in Eqs.(\ref{RG1}).
Mesoscopic values of $\Lambda_s$ and $c$ are related to
$K(\lambda)$ by
\begin{equation}
\Lambda_s (\lambda) = \frac{1}{\kappa (\pi K(\lambda))^2} \; , \;\;\;
c(\lambda) = \frac{1}{\pi \kappa K(\lambda)} \; .
\label{rel2}
\end{equation}
Since in the disordered case there takes place a renormalization of $c$, a
question may arise of what quantity $c$ should be used for the introduction of
the variable $y=c \tau$. For the derivation of Eqs.(\ref{RG2}) it
is natural to choose $c$ to be some small-scale value of $c(\lambda)$. For the
application of RG to a particular mesoscopic system the most appropriate
choice is $c=c(\lambda _*)$, where $\lambda _*$ is the large-scale cutoff
for $\lambda$, related to the size of the system. We notice, however, that
this difference leads only to small corrections in the parameter
$w^2 \ll 1$ to the RG equations, and thus may be neglected.

Finally, consider a special case of a regular system with a single impurity
(or any other defect). As it follows from the above analysis for regular and
disordered systems, in this case the effective action corresponds to
$XY$-model with the constraint that the vortex pairs be vertical and located
only in the vicinity of the line $x=x_0$, $\: x_0$ being the coordinate of the
defect. The free-energy-sign argument (entropy per vortex is
$\ln L + \mbox{\it finite term}$) immediately yields the critical value of
$K=1$,
in accordance with the result of Kane and Fisher \cite{Kane}. In the case of
one defect no renormalization of $K$ occurs. Physically it is clear since the
transition is associated only with the renormalization of the strength of the
defect: effective tunneling constant tends to zero at $K>1$ \cite{Kane}.

Now we use the effective action to consider the splitting
$\Delta_I$ of the SC level $I$ in a finite-size system as a function of $L$
and $K$. This splitting is due to the tunneling between the states
$\mid \! I \, \rangle$ and $\mid \! -I \, \rangle$ which means that the true
eigenstates of the finite-size system are superpositions of
$\mid \! I \, \rangle$ and $\mid \! -I \, \rangle$. To calculate
$\Delta_I$ {\it ab initio} one may take advantage of the conventional
instanton approach starting from the formula
\begin{equation}
\langle \, -I \mid e^{iHt} \mid I \, \rangle =
e^{-i(E_g + E_I)t} \sin (\Delta _I t)  \; ,
\label{f1}
\end{equation}
where $E_g$ is the groundstate energy, $E_I = 2\pi^2 I^2 \Lambda_s / L$ is the
supercurrent energy, $H$ is the system Hamiltonian. The next step is
to use the imaginary time $t=-i \beta$ in order to represent the l.h.s. of
(\ref{f1})
in terms of the effective action (\ref{S1}). Since the functional integral is
naturally introduced in terms of the coherent states, we represent $\mid \! I
\, \rangle$ as the expansion over the set of coherent states $\{ \mid \! \psi
\, \rangle \}$. Actually, we are interested only in the relation between the
expansion for $\mid \! I \, \rangle$ and that for the groundstate
$\mid \! 0 \, \rangle$.
In the homogeneous case this relation is trivial because of the Galilean
invariance: If
\begin{equation}
\mid 0 \, \rangle = \int D\psi \: Q[\psi] \mid \psi \, \rangle  \; ,
\label{f2}
\end{equation}
then
\begin{equation}
\mid I \, \rangle = \int D\psi \: Q[\psi]
\mid e^{i\frac{2\pi Ix}{L}} \psi \, \rangle  \; .
\label{f3}
\end{equation}
In the general case one may think of the state $\mid \! I \, \rangle$ as being
obtained from the homogeneous one (\ref{f3}) by semiadiabatic turning on the
external potential. The word "semiadiabatic" means that the turning-on time is
much greater than the inverse interlevel separation $\propto L$, but is much
less than $\Delta_I^{-1}$. Integrating over all the variables except for the
long-range phase field $\Phi$ we have
\begin{equation}
\langle \, -I \mid e^{-H\beta} \mid I \, \rangle =
A \int D\Phi \: e^{-S[\Phi ]}  \; ,
\label{f4}
\end{equation}
where $A$ is some independent of $L$ and $I$ normalization constant, the field
$\Phi (x,\tau )$ is determined at $0 \leq \tau \leq \beta$ and satisfies the
boundary conditions
\begin{equation}
\int_0^L dx \: \frac{\partial}{\partial x} \Phi (x,0) = 2\pi I \; , \;\;\;
\int_0^L dx \: \frac{\partial}{\partial x} \Phi (x,\beta ) = -2\pi I  \; .
\label{f5}
\end{equation}
To the first approximation, semiadiabatic turning on the external potential
does not affect Eq.(\ref{f4}), provided $\beta$ is much greater than the
turning-on time.

Calculation of the integral in (\ref{f4}) may be performed by a shift of the
variable $\Phi$:
\begin{equation}
\Phi = \Phi_1 + \Phi_0 \; ,
\label{shift}
\end{equation}
where the shifting field $\Phi_0$ is chosen to satisfy (below we turn to the
variable $y=c\tau$ and correspondingly $y_* = c\beta$ )
\begin{equation}
\triangle \Phi_0 = 0
\label{lapl}
\end{equation}
with the boundary conditions (\ref{f5}) ($\Phi \rightarrow \Phi_0$). Then
$S[\Phi ] = S[\Phi_1 ] + S[\Phi_0 ]$ and $\Phi_1$ may be integrated out taking
into account that
\begin{equation}
A \int D\Phi_1 \, e^{-S[\Phi_1 ]} =
\langle \, 0 \mid e^{-H\beta} \mid 0 \, \rangle = e^{-E_g \beta} \; .
\label{f6}
\end{equation}
As is easily seen, to meet the boundary conditions $\Phi_0$ should contain a
number of vortices (=instantons) with the total charge $2I$. So besides the
integration over $\Phi_1$ there should be also an integration over the
positions of the instantons: $D\Phi \rightarrow D\Phi_1 \,  \prod_{j=1}^{N_*}
dx_j dy_j$, $(x_j,y_j)$ being the position of the vortex $j$.
Generally speaking, there is also summation over the number of instantons,
$N_*$. For our purposes, however, we may confine ourselves to the case
$\beta \ll \Delta_I^{-1}$, where the main contribution comes from the $N_*$
corresponding to the minimal possible value of $S[\Phi_0]$. This implies that
all vortices are of the minimal possible charge of the same sign. We notice
also that in this case $\sin (-i \Delta_I \beta ) \approx -i \Delta_I \beta$.
Consequently, from (\ref{f1}) and (\ref{f6}) we obtain the relation for
$\Delta_I$:
\begin{equation}
\int \prod_{j=1}^{N_*} dx_j dy_j e^{-S[\Phi_0]} \: \propto \:
e^{-E_I \beta} \Delta_I \beta \; ,\;\;\;\;\;\; \Delta_I \beta \ll 1 \; .
\label{f7}
\end{equation}

Consider the most simple case of the splitting of the two-fold degenerate
groundstate level in the system with the gauge phase $\pi$, which we will
refer to as Aharonov-Bohm (AB) splitting, $\Delta_{AB}$. Upon the elimination
of the vector potential by the gauge transformation this case corresponds to
$I=\pm 1/2$. It is worthmentioning certain ring systems in which such a case
occurs automatically: fermions with
the even number of particles; lattice spins
with antiferromagnetic interaction and lattice bosons with the positive
hopping amplitude, both at the odd number of sites.

For $I=1/2$ minimal $S[\Phi_0]$ corresponds to $\Phi_0$ with the only one
vortex of the charge equal to unity:
\begin{equation}
S[\Phi_0] = i\gamma (x_0) + \frac{\pi^2 \Lambda_s}{2L}\beta + K^{-1} \ln L +
o(L^{-1}) \; ,
\label{S4}
\end{equation}
where $x_0$ is the $x$-coordinate of the vortex core. Taking into account that
\begin{equation}
\int_0^L dx_0 \: e^{i\gamma (x_0)} \: \propto \: \left\{
\begin{array}{ll}
\mbox{\it const} \;\;\;\;\;  \mbox{one defect} \\
\sim \sqrt{L} \;\;\;\;  \mbox{disordered system} \\
\;\;\;\;L \;\;\;\;\;\;\;  \mbox{commensurate with integer $\nu$} \\
\;\;\;\;0 \;\;\;\;\;\;\;\;\;  \mbox{otherwise}
\end{array} \right.
\label{f8}
\end{equation}
from (\ref{f7}) we obtain the estimate for the relative value of the AB
splitting.
\begin{equation}
\frac{\Delta_{AB}}{E_{AB}} \: \propto \: \left\{ \begin{array}{ll}
\;\;\;\;1/L^{1/K - 1} \;\;\;\;\;\; \mbox{one defect} \\
\sim 1/L^{1/K - 3/2} \;\;\;\; \mbox{disordered system} \\
\;\;\;\;1/L^{1/K - 2} \;\;\;\;\;\;  \mbox{commensurate with integer $\nu$} \\
\;\;\;\;0 \;\;\;\;\;\;\;\;\;\;\;\;\;\;\;  \mbox{otherwise}
\end{array} \right.
\label{Eab}
\end{equation}
Here $E_{AB} \equiv E_{I=1/2}$. Note that in the disordered case the
value of splitting is sensitive  to the particular realization of
disorder and experiences relative fluctuations of order unity from system
to system. Also, in the regular incommensurate case as well as in the
commensurate case with a fractional filling factor $\Delta_{AB}$ is equal to
zero because of the exact cancellation of contributions from instantons with
different $x_0$'s.

For disordered and commensurate systems the laws (\ref{Eab}), generally
speaking, take place only at sufficiently large $L$, where $K$ reaches its
macroscopic limit. So they need be revised in the vicinity of the critical
points, where renormalization of $K$ is essential and, moreover, is very slow
with respect to $L$. (Such a problem does
not emerge in the single-impurity case,
since there is no renormalization of $K$.) To allow for the renormalization we
notice that the only place in the effective action (\ref{S4}) where smaller
scales of distance may be important is the vortex core energy $E_V(L) = K^{-1}
\ln L$, which now is to be estimated more accurately. This may be done just by
noting that according to the above-mentioned meaning of the
variable $w$ in RG equations it is related to $E_V$ by
\begin{equation}
w^2(L) \propto \left\{ \begin{array}{ll}
L^3 e^{-2 E_V(L)} \;\;\;\; \mbox{disordered} \\
L^4 e^{-2 E_V(L)} \;\;\;\; \mbox{commensurate}
\end{array} \right.
\label{w}
\end{equation}
Thus we obtain
\begin{equation}
\frac{\Delta_{AB}}{E_{AB}} \propto w(L) \; ,
\label{RG_Eab}
\end{equation}
where $w(L)$ should be taken from RG equations. In particular, at the critical
point, where both (\ref{RG1}) and (\ref{RG2}) yield
$w(L) \sim 1/ \ln L$, we have
\begin{equation}
\frac{\Delta_{AB}}{E_{AB}} \propto 1/ \ln L  \;\;\;\;\;\;\;
\mbox{(at the critical point)} \; .
\label{Eab_crit}
\end{equation}

In the case of commensurate system with a fractional $\nu \propto 1/p$ only
the instantons whose charges are multiples of $p$ are relevant. Clearly, this
leads to the fact that only some special SC levels are split. Namely, the
levels with $I$'s being the multiples of $p/2$. Consequently, at an odd $p$
split SC levels arise only at the gauge phase $\pi$, while at an even $p$
the gauge phase $\pi$ renders all SC levels degenerate. Estimation of the
splitting of the first split SC level ($I=\pm p/2$) is done in the
complete analogy to the estimation of $\Delta_{AB}$. The only difference is
that the charge of the instanton now is $p$ rather than unity. So we readily
get
\begin{equation}
\frac{\Delta_{p/2}}{E_{p/2}} \: \propto
1/L^{p^2/K - 2} \;\;\;\;\; \mbox{commensurate with $\nu \propto 1/p$} \; .
\label{E_p_2}
\end{equation}
This result is a generalization of Eq.(\ref{Eab}) for the commensurate case.
Obviously, Eqs.(\ref{RG_Eab},\ref{Eab_crit}) (with AB$\; \rightarrow p/2$)
are also valid.

If we try to estimate by the same technique the splittings of higher SC
levels, when more than one instanton is necessary, we have to deal in
(\ref{f7}) with a divergency of the
integration with respect to $y_j$'s. This is a standard problem one faces when
considers in imaginary time the tunneling between the excited states rather
than between ground-state ones. Instead of thinking of a trick of
circumventing this difficulty, we prefer to turn to another language of
description. In the next section we introduce the effective Hamiltonian, whose
relation to the effective action (\ref{S1}) is clearly traceable, and which
allows to treat the problem of the SC splitting as well as the hybridization
with phonons most naturally.

\section{Effective-Hamiltonian approach}
\label{sec:3}

In this section  we demonstrate how one may
construct an effective Hamiltonian for the SC dynamics and calculate
the  ratios $\Delta_I /\Delta_{AB}$ as well as the resonances with the
phonons.

In the Fourier-transformed fields the Hamiltonian of the superfluid phase
can be written as
\begin{equation}
H_0=\sum_{q} \omega_q\;b^{\dag}_q\:b_q \;\;+{\pi \over 2L}(v_N\: \delta N^2+
4 v_I I^2)\; ,
\label{H1}
\end{equation}
where
$v_I = \pi \Lambda_s$, $v_N = (\pi \kappa)^{-1}$,
$b^{\dag}_q $ is the standard Bose creation operator,
 $\omega_q \approx c \: \mid \: q\: \mid \: $ in the long-wavelength limit.
For completeness, we have introduced the term
associated with the energy change when we add particles to the system,
$\delta N=N-N_0$.

We start by noting that the instanton is nothing else but an effective
vertex connecting nearest SC. Thus in the most general form we may
present the instanton at the point $x$ as
\begin{equation}
H_{int}(x) =\sum_{I} \left( {V_{\Lambda } \over L}  e^{i \gamma (x) }\:
\hat{O}(b )
\: c^{\dag}_{I+1} c_I \:+\: h.c. \right) \; ,
\label{V1}
\end{equation}
Here $c^{\dag}_{I+1}$ is the raising operator of the quantum number $I$,
$\gamma (x)$ is given by Eq.(\ref{gamma1}),
 and $\hat{O}$ is an operator acting on the phonon states.
The dependence on the coordinate comes from the fact that
an intrinsic momentum of the current state in the homogeneous system
equals to $I 2 \pi n$
and the Hamiltonian (\ref{V1}) changes $I$ by $1$. Indeed,
the SC is obtained by shifting all the particles in the momentum
space  by $I$ quanta with the total momentum change $NI\; 2\pi /L$.
In disordered systems the particle density is no longer
homogeneous and we must replace the average particle density
with the local one $n \to n_0(x)$ according to (\ref{gamma1}).
To satisfy the RG equation for the effective vertex
$w$ (see also below)
the scaling dimension of  $\hat{O}$ should equal to $1-1/K$.

As shown by Haldane \cite{Hald81a} (and will be demonstrated
explicitly below for the particular case of electron backscattering
from the impurity) the bosonic field responsible for the current
relaxation is
\begin{equation}
\theta (x)=(\theta_0 + \gamma (x)) + i  {1 \over \sqrt{K} }
\sum_{q\ne 0} \sqrt{{ 2 \pi \over \mid q \mid L }}
{\rm s i  g n} (q)
(b_{q} + b_{-q}^{\dag})e^{-iqx} \;.
\label{theta}
\end{equation}
and the original particle field is given in terms of exponential
functions of $\theta (x)$. By noting that the phase $\theta_0$
is the canonical momentum for $I$, the resulting
effective Hamiltonian may be rewritten as
\begin{equation}
H_{int}(x)=\sum_{-\pi < \theta_0 \le \pi} \left( {V_{\Lambda } \over L}
  e^{i (\theta_0 + \gamma (x))}\: \hat{O}(b )
\: c^{\dag}_{\theta_0} c_{\theta_0} \:+\: h.c. \right) \; ,
\label{mrep}
\end{equation}
Clearly, the first two terms in the exponent are those in the
definition of $\theta (x)$ given above, and we identify the operator part
of the vertex $\hat{O}$ in Eq.(\ref{V1}) with the $q \ne 0$ components
of the phase field  $\theta (x)$
\begin{equation}
 \hat{O}(b ) = {L \over \Lambda}
  \exp \{ i \theta_{q \ne 0}^{(\Lambda )}(x) \}  \; ,
\label{O}
\end{equation}
where the first factor is necessary for the correct scaling of
$\hat{O}$.
The sum over momentum has an upper cutoff at some scale $q \le 2\pi /\Lambda$
where one has to define the "bare" amplitude $V_{\Lambda}$
which  depends on the
microscopic parameters of the initial Hamiltonian and the cutoff.
One can not obtain the exact value of the parameter
$V_{\Lambda}$ from this analysis (except for the scaling behaviour
mentioned above). The other way of deriving the effective Hamiltonian
mathematically rigorously
is by mapping the effective Lagrangian Eqs.(\ref{S1},\ref{K}) to the
equivalent sine-Gordon model \cite{Col}, which is in fact done
by Eq.(\ref{mrep}) since $H_{int} \sim \cos \theta (x)$.

There is yet another simple microscopic derivation of Eq.(\ref{O}) for
the particular case of electron backscattering from the impurity potential.
In the bosonic representation the original right- and left-moving electron
fields (labeled by the index $p= \pm 1$) have the form \cite{Hald81b}
\begin{equation}
\Psi_p^{\dag} (x)={1\over \sqrt{\Lambda}}{\bf U}_p e^{-ipk_Fx}
\exp \left\{  \sum_q \sqrt{{2\pi \over  \mid \: q \: \mid L }}
{\rm s i  g n} (q) \:
C_{pq}(b_q+b_{-q}^{\dag} )e^{-iqx} \right\} \;,
\label{Psi}
\end{equation}
\[
C_{pq}= \left\{
\begin{array}{ll}
\cosh (K) \;,\;\;\;\;pq>0 \\
\sinh (K) \;,\;\;\;\;pq<0
\end{array} \right.
\]
where the Fermi momentum  $k_F$ equals to $\pi N/L$, and
${\bf U}_p$ is the ladder operator increasing the number of $p$-particles
by one. If we substitute
these expressions into the backscattering Hamiltonian, $H_B =
V_B \Psi_r^{\dag} \Psi_l + h.c.$, which is responsible for the
current relaxation on impurities,
we immediately arrive at the effective Hamiltonian
Eq.(\ref{mrep},\ref{O}) \cite{GP}. An important new point here is that
it is valid for any 1D system in the superfluid phase,
provided one uses the proper definition of the
quantum number $I$, which has the particular  meaning of
the difference between the number of
left- and right-moving particles for fermions only.

A short physical discussion is in order here. One may consider the
effective Hamiltonian (\ref{V1}) as resulting from the coupling
between the SC states and sound waves of the form
\begin{equation}
\sim  c^{\dag}_{I} c_I {Ic \over \sqrt{K} }
\sum_{q\ne 0} \sqrt{{ 2 \pi  \mid q \mid \over  L }}
{\rm s i  g n} (q)
(b_{q} - b_{-q}^{\dag})e^{-iqx} \;.
\end{equation}
As is well known, this interaction can be eliminated by the standard
on-site polaronic transformation in expense of the additional
operator structure $\hat{O}$  in the hopping term. Thus any
SC state causes a long-range deformation in the phonon
field, and this deformation is growing linearly with the
quantum number $I$. In the fermion system this is equivalent to
"dressing" each extra left/right-moving particle by density
fluctuations. Now any attempt to change the SC state will result
in the readjustment of the phonon vacuum to the new value of $I$. It turns out
that the dominant contribution to the effective low-energy vertex
is due to this reconstruction of density fluctuations rather then
due to the "bare" process of current relaxation which defines the
high-energy prefactor here. As $K$ goes to zero the coupling between
the SC state and density fluctuations diverges. This
limit describes the case of weakly interacting bosons and quasi
1D superconducting rings (see section \ref{sec:5}), where even the
lowest SC state (which still has a finite momentum!)
involves {\it large} number of particles with {\it small} momenta
 moving in the same direction. Now the process of current relaxation
is through the backscattering of many particles simultaneously,
rendering the low-energy vertex extremely small. Thus a one-particle
description of SC in terms of the number of left- and right-moving
carriers is clearly misleading for $K \to 0$; the relevant
language in all cases being that based on the  quantum number $I$.

One may now straightforwardly check that the renormalization group
equations in the disordered case  are correctly described by the effective
Hamiltonian.
The RG equation for the vertex is obtained by a simple
averaging of the operator $\hat{O}$ over the momentum range
$2\pi /\Lambda^\prime < \mid q \mid <2\pi /\Lambda$
where $\Lambda^\prime = \Lambda + d\Lambda $.
In the second order perturbation theory
the renormalization of the superfluid energy is due to the
virtual transition of the system to the neighbouring current state
with the simultaneous emission of the sound mode with
$2 \pi /\mid q \mid \in  (\Lambda , \Lambda^\prime )$
\begin{equation}
\delta E_I =-(V/L)^2 {1 \over K} 2\int_{ (\Lambda, \Lambda^\prime ) }
{dq \over q}
\bigg[ {1 \over 2\pi c/\Lambda +E_{I+1}-E_{I} } +
{1 \over 2\pi c/\Lambda +E_{I-1}-E_{I} } \bigg] \;.
\label{EI}
\end{equation}
where $E_I=2\pi v_I I^2/L$. By expanding the denominators up to the
second order in 1/L we find the RG equation for the current velocity:
\begin{equation}
{d\ln v_I \over d \ln \Lambda }=
- { 4 V^2 \Lambda \over \pi^2 c^2K^2 L}\;.
\label{vI}
\end{equation}
The renormalization of the sound velocity is due to the virtual
transitions from $I$ to $I \pm 1$ with the simultaneous absorption
of the sound wave and emission of the high-frequency mode. In full
analogy with the previous calculation we have
\begin{equation}
{d\ln c \over d \ln \Lambda }=
- { 2 V^2 \Lambda \over \pi^2 c^2K^2 L}\;.
\label{c}
\end{equation}
As expected, one may neglect the renormalization of $v_I$ and $c$
in the superfluid phase when dealing with the single impurity scattering
(the r.h.s. of Eqs.(\ref{vI},\ref{c}) is $1/L$ in this case).

For disordered potential the effective Hamiltonian is an obvious sum over
the impurity positions, and in Eq.(\ref{V1}) one has to replace
$V \to \sum_{x_i} V e^{i \gamma ( x_i ) } \sim V (L/\Lambda )^{1/2} $.
Now the scaling equation for the vertex is changed to
$dV/d \ln \Lambda = V (3/2 - 1/K)$, in agreement with Eq.(\ref{RG2}).
Correspondingly, in Eqs.(\ref{vI},\ref{c})  one has to replace
$V^2$ by $V^2 L/\Lambda $. By
introducing the dimensionless coupling parameter
$w=2V/\pi c K$
we rewrite the RG equations as
\begin{eqnarray}
{d\ln v_I \over d \ln \Lambda }& = &
- w^2 \;,\nonumber \\
{d\ln c \over d \ln \Lambda }&=&
- w^2/2  \;,
\label{vIc}
\end{eqnarray}
or, using the universal relations $c=v_IK=v_N/K$,
\begin{equation}
{dK \over d \ln \Lambda } = w^2 K \;; \;\;\;\;
v_N=const \;.
\label{KvN}
\end{equation}
Thus we have reproduced the RG equation (\ref{RG2}) of section \ref{sec:2}.

In exactly the same manner one can construct an effective Hamiltonian
for the commensurate system. Let $n =1/p$. Then the vertex connecting
the states $I$ and $I\pm p$ is defined by
\begin{equation}
H_{comm} =\sum_{I} \int {dx \over \Lambda } \left(
{ V \over \Lambda }   e^{ip\theta_{q\ne 0}(x) }
\: c^{\dag}_{I+p} c_I \:+\: h.c. \right) \; .
\label{Hc}
\end{equation}
Due to the translational symmetry of the commensurate problem there is
no renormalization of the sound velocity. For the same reason
only states with zero total momentum are hybridized with the
topological excitations and e.g. the
correction to the superfluid energy is defined by virtual transitions
to the state with two high-frequency modes $q_1=-q_2 \in
(\Lambda , \Lambda^\prime )$. Now
\begin{equation}
{dK \over d \ln \Lambda } =  p^6/16\:w^2 \;; \;\;\;\;
c=const \;.
\label{Kc}
\end{equation}
The coefficient in front of the $w^2$ term
can be absorbed into the redefinition of $w$ (cf. Eq.(\ref{RG1})).

The effective-Hamiltonian approach reduces the discussion of the
superfluid - insulator transition to that of a "particle" localization
at a given site $I$. In the ground state (when $\Lambda =L$) the
current Hamiltonian is identical to the tight-binding model with the
hopping amplitude $V_L /L$ in the static external potential given
by the supercurrent energy
\begin{equation}
H(L) \approx \sum_{I} {2\pi v_I  \over L } (I-\phi /\phi_0 )^2 \;
c^{\dag}_{I} c_I +
\sum_{I} \left( {V_{L} \over L} \: c^{\dag}_{I+1} c_I \:+\: h.c. \right) \;.
\label{HL}
\end{equation}
If under scaling one finds $V_L \ll v_I$, then the potential energy
prevents the current state from delocalization, and we have perfect
metallic conductance. Only at some special values of the magnetic flux
 through the ring (e.g. integer or half-integer flux), when the states
$I$ and $-I$ are in resonance, the current is delocalized between the
two states, but the energy splitting of the resonance is much less than
$2\pi v_I /L$.

The superfluid-insulator transition is described as a delocalization
of the current state, when $V_L \gg v_I$. In this regime the
topological invariant is delocalized in a band and $I$ is no longer
a well-defined quantum number \cite{GP}.

Let us now calculate  the splittings between the resonating states.
If we neglect the hybridization with
phonons and work within the Hamiltonian (\ref{HL}), then the splitting
is given by the perturbation theory of $2I$-th order in $V_L$
\begin{eqnarray}
\Delta_I & =& 2{V_L^{2I} \over L} \prod_{l=1-I}^{I-1} {1 \over
2\pi v_I (I^2 -l^2) } \nonumber \\
& =& {4\pi v_I \over L} \: \left( {V_L \over 2\pi v_I} \right) ^{2I}\:
{1 \over [(2I-1)!]^2} \;.
\label{DI}
\end{eqnarray}
If the ring is in the flux $\phi = \phi_0 /2$ then the splitting between the
states  $I$ and $1-I$ is described by the same expression (\ref{DI}) with
the replacement $I \to I-1/2 $.

In deriving Eq.(\ref{DI}) we considered only the path from $I$ to $-I$
which corresponds to the ground state of the bosonic field in
{\it all} intermediate states. In the general case one has to
consider the possibility of exciting/absorbing
an arbitrary number of sound waves in any intermediate state. To this end we
slightly "undress" the ground-state vertex and write the final low-energy
Hamiltonian in the form
\begin{equation}
H_{eff} = \sum_{I} {2\pi v_I  \over L } (I-\phi /\phi_0 )^2 \;
c^{\dag}_{I} c_I +
\sum_{I}\left(  {V_{L} \over L}\:{e^{i\theta }
\over \langle e^{i\theta }\rangle }
 \: c^{\dag}_{I+1} c_I \:+\: h.c. \right) \;.
\label{HLE}
\end{equation}
which is cutoff independent. First, let us consider the case of single
impurity, when the spatial dependence of $V_L$ is $\delta (x-x_i)$ at the
impurity position $x_i$.
Using standard properties of bosonic
operators we readily calculate all the matrix elements
\begin{eqnarray}
& & \langle \{ N_q \} \mid \; {e^{i\theta } \over \langle
e^{i\theta }\rangle } \; \mid
 \{ N_q+S_q \} \rangle =
\prod_q r(N_q,N_q+S_q, {2\pi \over LK \mid q \mid }) \;, \nonumber \\
& & r(N,N+S,x)={e^x x^{S/2} \over \sqrt{N! (N+S)!} }
\: \sum_{n=0}^{\infty} {(-1)^n (N+S+n)! \over x^n n! (n+S)!}
 \left\{
\begin{array}{ll}
1  \;\;\;\;\;\;\;\;\;\;\;S \ge 0 \\
(-1)^S \;\;\;S<0
\end{array} \right. \;.
\label{ME}
\end{eqnarray}
In principle, the problem of the low-energy spectrum is solved now, because
we know explicitly all the matrix elements. In a large system the coupling
$V_L$ is macroscopically small, and apart from extremely narrow
resonances at $1/K=$ integer, when one has to solve a finite-size
secular equation, it is sufficient to restrict
ourselves to the perturbation theory in $V_L$. Below we calculate
analytically the energy splitting of the first SC state.

First, we note that the coupling to phonons does not modify the splitting
of the ground state in a ring with $\phi = \phi_0 /2$, simply
because there are no intermediate states in this case (see Eq.(\ref{Eab}))
\begin{equation}
\Delta_{AB} = {4\pi v_I \over L} (V_L /2\pi v_I ) \;.
\label{D0}
\end{equation}
With the coupling to phonons taken into account
 the expression for the first SC splitting acquires the form
\begin{equation}
\Delta_1 = \Delta_{AB}^2 {L \over 4\pi v_I }
\sum_{ \{ q,S_q \} } { \prod_q (-1)^{S_q} \mid
r(0,S_q,{2\pi \over LK\mid q \mid }) \mid ^2 \over
1-\sum_q {LK\mid q \mid \over 2\pi } S_q } \;.
\label{DP}
\end{equation}
Next we introduce the integer variable $M$ and
write down the spectral representation for the $\delta$-symbol
$\delta (M-\sum_q (L\mid q \mid/ 2\pi )S_q )$. This allows to
calculate the contribution of each phonon mode separately
\begin{eqnarray}
R={4 \pi v_I \Delta_1 \over L\Delta_{AB}^2 }& = &
\int_{-\pi }^{\pi} {dt \over 2\pi } \sum_{M=0}^{\infty}
{e^{-iMt} \over 1-KM} \exp \left\{
-{2 \over K} \sum_{p=1}^{\infty} {e^{ipt} \over p } \right\}
\nonumber \\
& =& \sum_{M=0}^{\infty} \int_{-\pi }^{\pi} {dt \over 2\pi }
{e^{-iMt} (1-e^{it})^{2/K}  \over 1-KM} \;.
\label{R}
\end{eqnarray}
Straightforward evaluation of this expression gives the final answer
\begin{equation}
R={\pi /K \over \sin (\pi /K ) }\: {\Gamma (1+2/K) \over
\Gamma^2 (1+1/K) } \;.
\label{Rf}
\end{equation}
Note that $R=1$ if we neglect the  hybridization with phonons.

Thus we found that for $1/K=$ integer the splitting diverges due to
the resonance between the SC and phonons. For small $V_L$ this
divergence is accurately described by Eqs.(\ref{R},\ref{Rf})
except for a tiny region of order $V_L /v_I$ around integer values of $1/K$.
However in this region the ratio $R$ is already very large, and the
dominant contribution comes from the coupling to a small group of
phonon states with the same energy $\approx 2\pi v_I /L$.
For the particular case of $1/K \approx 2 $, which will be compared in the
next section with the exact numerical diagonalization,
one has to take into account the states $\mid N_{q} = 2\rangle , \;
\mid N_{-q} = 2\rangle ,\; \mid N_{q} = 1,N_{-q} = 1\rangle ,\;
\mid N_{2q} = 1\rangle ,\;$ and $ \mid N_{-2q} = 2\rangle $ where $q=2\pi/L$.
The hybridization matrix elements
between these states and SC squared
are $2,\;2,\;4,\;1,\;$ and $1$, correspondingly.
By writing a secular equation for this group of levels we obtain a complete
picture of level crossings between the SC and sound modes.

Quite generally, at the point of resonance {\it all} energy splittings
are of the same order $\Delta_I \sim \Delta_{AB}$. Indeed, for
integer $1/K$ the instanton is always between the states with the same
energy (for the combined system SC $+$ phonons),
so the splitting is given by the amplitude of a single
transition $\Delta_{AB}$ no matter how many instantons are required to connect
the resonating SC states.

This technique may be applied for calculating the energy splittings
in commensurate systems as well. An important  difference with the
impurity  case is that now
the supercurrent  energy change is
$\delta E_{I- p,I}=2\pi v_I (2pI-p^2)/L$ and resonating
states are given by $I=p/2 \pm mp$,
and  only sound waves with the zero total
momentum are hybridized with the SC.
We find that for commensurate systems
we always have a resonance between the SC and sound at the point of
superfluid-insulator transition. Indeed, for $K=p^2/2$ the energy change
$\delta E_{I\pm p,I}=2\pi c/L \;2(2n-1)$ is given by even number of
the phonon energy quanta.

Strictly speaking, there is only one cutoff-dependent parameter $V_L$
(apart from the Luttinger Liquid parameters $c$ and $K$) which
controls the low-energy spectrum of the system. Once this parameter
is fixed, say from the $\Delta_{AB}$ splitting, all higher energy splittings
and level crossings can be found explicitly.
However in the disordered system we face a problem. By construction the
low-energy vertex $V_L(x)$ in this case is a random function of coordinate,
and depends on the particular disorder realization in the sample.
Now, even if we define the mean value of $<V_L>$ from the lowest
splitting $\Delta_{AB}$, we still are uncertain about the integrals
$\int_0^L dx e^{\pm i2\pi mx/L} V_L (x) $ which define the hybridization
matrix elements with the phonon modes and take care about
momentum non-conservation in disordered system. The only thing which is
certain is that for the lowest levels with $m \sim O(1)$
these integrals have random values of order $<V_L>$ in magnitude.
It makes the direct comparison between the theoretical calculation
(\ref{Rf}) and the numerical result for $R$ inconclusive, unless
additional parameters are introduced into the theory to account for the
random variations of the above-mentioned integrals (see more in the
discussion  of the numerical results in section \ref{sec:4}).

The above discussion dealt with the properties of an isolated 1D ring.
Obviously, by  considering the lowest energy levels we lack any kind of
dissipation in the system. In the ideal picture the current state $I$
obtained, e.g., by nonadiabatic switching on the magnetic flux through
the ring will persist forever, or demonstrate undamped
oscillations with the frequency $\Delta_I$. The most effective dissipation
mechanism is due to the coupling between the ring and substrate. One has to
take an insulating substrate in order to eliminate the conduction electrons
which form a thermal bath with the largest low-frequency density of states,
and thus are the most dangerous source of decoherence. To couple the SC states
to phonons of the substrate
we have to consider the variation of $H_{eff}$ originating from
the lattice long wave-length distortion $u(x)$. The effect of $u(x)$ is
two-fold:
to change the  kinetic  and  potential energy. For example, the
distortion with the wave-length of order $L$ (which can absorb the energy
difference $\delta E_{I,I-p}$ )
may influence the total length of the ring $\delta L
\sim Lu$ and result in the $u$-dependent terms both in the kinetic and
potential energy. For disordered sample the positions of impurities
$\{ x_j \} $  depend on the substrate deformation giving
$\delta V_L \sim V_L u$.
One may also take into account the dependence of $K$
(or the particle-particle interaction) on the inter-particle separation.

It is easy to check that both the nondiagonal and diagonal couplings give the
same value of the relaxation time $\tau$. We write the diagonal interaction
Hamiltonian as
\begin{equation}
H_{diss} \sim \alpha {v_I \over L}u \sim
\alpha {v_I \over L} \left[ \sum_k \sqrt{k/Ms} (d_k+d^{\dag}_k ) \right]
 \;,
\label{HD}
\end{equation}
where $d^{\dag}_k $ are the bulk phonon creation operators, $\alpha $
is the dimensionless parameter depending on the microscopic
details and geometry  of the particular system, $M$ is the substrate ion mass,
and $s$ is the bulk sound velocity. Now, using standard results for the
relaxation time  in a double-well system coupled to phonons (see e.g.
\cite{KM}) we find
\begin{equation}
\tau^{-1} \sim \left( {V_L \alpha v_I \over L^2 } \right) ^2
{ \mid E_{I,I-p} \mid \over \rho s^5 } \sim
{V_L^2 \alpha^2 v_I^3 \over \rho s^5 L^5 }\;, \;\;\;(T=0) \;.
\label{tau}
\end{equation}
Here $\rho $ is the substrate density. If we couple the substrate to the
nondiagonal terms in Eq.(\ref{HLE}) by expanding
\[
V_L/L \to V_L/L \:(1+ \alpha^\prime u )\; ,
\]
then the final result for $\tau^{-1}$ coincides with the Eq(\ref{tau}).

The relaxation rate should be much less  than the level splitting
$\Delta_I$ if we want to observe the coherent oscillations. In the mesoscopic
limit we find $\Delta_I \tau \sim V_L^{2(I-1)} L^4$. Substituting here the
scaling behaviour of $V_L$, we obtain the criteria defining which levels
will demonstrate coherence  at $T=0$
\begin{equation}
 I <  \left\{
\begin{array}{ll}
(1+K)/(1-K)   \;\;\;\;\;\;\;\; \mbox{single impurity} \\
(2+K)/(2-3K) \;\;\;\;\;\; \mbox{disordered}\\
p^2/(p^2-2K) \;\;\;\;\;\;\;\;\;\;\; \mbox{commensurate}
\end{array} \right. \;.
\label{ICR}
\end{equation}
We see that the broadening of the two lowest doublets $\Delta_{AB}$ and
$\Delta_{1}$ is always small. Very close to the superfluid-insulator
transition
we can safely neglect dissipation for all levels. To get a better feeling
of how long the relaxation time could be in a real system, we estimate
$\tau^{-1}$ from Eq.(\ref{tau}) using $\gamma \sim 1,\; v_I\sim s,\;
L\sim 10^4 A^o,\;$ and the maximum possible $V_L \sim v_I$. For this
particular
choice of parameters  $\tau \sim \rho L^5 \sim 1\:$week.

\section{Numerical spectroscopy}
\label{sec:4}

In this section we report exact numerical spectroscopic results for 1D
finite-size boson Hubbard models. The Hamiltonians we examine can be written
in the most general form as (all the parameters are scaled in the units of the
hopping amplitude)
\begin{equation}
H = \sum_{i=1}^{N_a} \left\{ -(a_i^+ a_{i+1} + \mbox{h.c.}) + \epsilon_i n_i +
{U \over 2} n_i (n_i -1) + V n_i n_{i+1} \right\} \; .
\label{Hub}
\end{equation}
Here $a_i$ is the annihilation operator on the site $i$; $n_i = a_i^+ a_i$,
and $\{ \epsilon_i \}$ are random numbers uniformly distributed over the range
from $-W/2$ to $W/2$.
(In the single-impurity case all the $\epsilon_i$'s are
equal to zero except for $\epsilon_1$.) The ring topology implies that the
sum of the site
indices is understood modulo $N_a$, the total number of sites. The gauge phase
$\pi$, if necessary, is introduced by changing the sign of one of the hopping
amplitudes (anti-periodic boundary condition).

For the calculation of the scalings of SC splittings with increasing $N_a$ we
use hard-core models ($U=\infty$) which at a given $N_a$ have much less
Hilbert-space dimension as compared to the full models ($U \neq \infty$), and
thus allow consideration of larger systems. In this case the term with $V$ is
necessary to control the parameter $K$. In the case of a full model, where
this term does not introduce any qualitative difference with respect to the
on-site term $U$, we set $V=0$.

Our diagonalization procedure \cite{KPS} is arranged as follows. First,
starting from some trial wavefunction we calculate  a certain number of the
lowest energy levels and corresponding wavefunctions by modified Lanczos
method \cite{Dag}. The set of approximate eigenfunctions is reconstructed
from Relay's tridiagonal matrix \cite{Pis}, and the trial wavefunction is
expanded in it. As is known, the set inevitably involves a substantial number
of spurious states, due to numerical errors. These states, however, may be
easily identified by their negligible contribution to the expansion of the
trial wavefunction. Upon exclusion of the spurious states the set is
subjected to the orthogonalization and correction by Newton method. The
relative (with respect to a characteristic interlevel spacing) errors of the
energy level calculation are typically of order
$10^{-13} \div 10^{-11}$ for the
groundstate, and of order $10^{-9} \div 10^{-5}$ for some ten first
excited states. The problem of degeneracy is solved by repeating the procedure
with a new trial function, chosen to be orthogonal to all previously
obtained eigenstates. The combined set of eigenfunctions is again
orthogonalized and corrected by Newton method.

In the case of the regular system we take advantage of the translational
symmetry to proceed separately for each momentum sector.

Now we turn to the results. To get an idea of how the spectra look like in
this or that case, we show the pictures of spectrum evolution with increasing
on-site disorder (Fig.\ 1) and on-site interaction (at a fixed disorder)
(Fig.\ 2), and as a function of the nearest-neighbor interaction in the case
of commensurate hard-core model (Fig.\ 3). All the energies are reckoned from
the ground states. Besides, in the case of disordered system we found it
convenient to normalize the energy values in such a way that the middle of the
first phonon doublet always corresponds to unity. With this normalization the
value of the first SC coincides with $K^{-1}$.
The level identification is performed in accordance with \cite{KPS},
where the numerical spectra for a disordered 1D ring were reported for the
first time.

Though it is not directly related to the subject of the present paper, it is
interesting to discuss the behavior of the hard-core model at large negative
$V$'s. From general considerations it is clear that at a certain negative
$V=V_*$ there should occur an ordinary condensation into a macroscopic drop.
So that at $V<V_*$ the first $N_a -1$ energy levels (associated with the
motion of the drop as a whole with momenta $=1,2,3, \ldots , (Na-1)$) should
lie very close to the ground-state one (corresponding to the zero-momentum
motion of the drop). It is precisely what is seen in Fig.3.
{}From the symmetry
of the hard-core model (isomorphism with the spin-$1/2$ system) it immediately
follows that in our case $V_* =-2$ is the point where, speaking the spin
language (isotropic ferromagnetic Heisenberg Hamiltonian), the transition
to the ferromagnetic ground state takes place. To minimize the energy
below this point a system with fixed $z$-projection of the full spin (fixed
number of particles in our case) would desire to be divided into two parts:
with spins up and spins down, respectively (a part with empty sites and a part
with completely occupied ones). In a regular system this symmetry breaking is
hidden by translation invariance, but may be revealed by calculating some
relevant correlator. Consider e.g. $Q_3 = \langle n_{i_1} n_{i_2} n_{i_3}
\rangle / \langle n_{i_1} \rangle ^3$, where $\mid i_1 - i_2 \mid =
\mid i_2 - i_3 \mid = \mid i_3 - i_1 \mid = N_a /3$. In the macroscopic limit
we have $Q_3 =1$ for homogeneous liquid and $Q_3 \rightarrow 0$ for a
condensed one. Our calculations for $N_a =18$ with the filling factor
$\nu =1/2$ yield $Q_3 \approx 0.90$ at $V=-1.8$ and $Q_3 \approx 0.14$ at
$V=-2.3$.

In connection with the transition at $V=-2$ in the hard-core model it is
worthmentioning that just at this point there exists an analytical
Bethe-ansatz treatment recently reported by Sutherland \cite{Suth}, which
reveals the appearance of macroscopic bound complexes.

The scaling of the single-impurity AB-splitting (Fig.\ 4) is in excellent
agreement with the macroscopic theory. Here and below the dashed line
("theory") is plotted in such a way that it passes through the last
(maximum $N_a$) point and has a slope corresponding to Eqs.(\ref{Eab},
\ref{E_p_2}), the value of $K$ being obtained from the numerical spectrum at
this point (sound velocity can be taken from the first phonon level while
$\Lambda_s$ is available e.g. from the first SC).

The analogous results for the disordered system (Fig.\ 5) demonstrate a good
agreement with the theory only at larger $N_a$. In fact, this is what one
might expect to get bearing in mind that the value of $K$ we deal with
($\sim 0.5$) is not very far from the critical one ($=2/3$), so that the
renormalization of $K$ due to the disorder may be significant. In accordance
with Eq.(\ref{Eab}) the decrease of $\Delta_{AB}$ with increasing $N_a$ at
$K \sim 0.5$ is rather slow. Hence, to reveal it against the statistical
dispersion, characteristic of the disordered case, an extensive averaging is
required.

Fig.\ 6 shows the scaling of $\Delta_1$ in the commensurate case. The results
at $V=-0.5$ are in excellent agreement with Eq.(\ref{E_p_2}), while at
$V=-1.5$ there is a slight deviation (which, however, does not exceed $15$\%).
Since this deviation has a tendency to decrease with increasing $N_a$, it may
be accounted for as a small-size effect.

Fig.\ 7 shows the scaling of $\Delta_1$ for the commensurate system at the
critical point and demonstrates the excellent agreement with the prediction of
the macroscopic theory that the plot will approach a straight line as
$N_a \rightarrow \infty$. It should be noted that for the half-filled 1D
hard-core model the critical point (corresponding to $V=2$ where the model is
isomorphic to the isotropic Heisenberg antiferromagnetic spin-$1/2$ chain) was
extensively studied \cite{Aff_et_al}. In particular, it was shown that some
excited levels demonstrate logarithmic ($\sim 1/ \ln N_a$) relative shifts of
their positions as compared to the macroscopic limit. Speaking superfluid
language \cite{we}, we understand that these are the states that have either
non-zero topological quantum number $I$, or extra particles/holes. (Purely
phonon excitations can not demonstrate such shifts since the sound velocity
is irrenormalizable quantity.)

To verify the correctness of the effective-Hamiltonian approach
Eqs.(\ref{V1},\ref{O}) we calculate numerically the relation between the
SC splittings $\Delta_1$ and $\Delta_{AB}$ for the disordered ($W=0.25$)
system of 7 bosons on 11 sites. The coupling parameter $U$ varied in the
range $2.6 \le U \le 6.2$ where the first SC level crosses the group
of phonon levels with energies $\approx 4\pi c/L$. The numerical ratio
$R=2E_1\Delta_1/\Delta_{AB}$ is shown in Fig.8.
Unfortunately, the direct comparison with the theoretical expression
(\ref{Rf}) is not very conclusive here because it was calculated
by assuming that {\it all} 5 phonon levels with energies
$\approx 4\pi c/L$ are degenerate. However in a very small ring this is
not the case even approximately as is clearly seen in Fig.2. The single
resonance is split and in the parameter range $U <3$ other
multi-phonon states (also strongly split) practically wash out any
sign of the phonon spectrum quantization  in multiples of $2\pi c/L$.
Under these conditions one may not expect to see any reasonable agreement
between the calculated ratio  Eq.(\ref{Rf}) and the numerical
results on such small systems.

One may try to improve the theory by adjusting the calculated ratio
(\ref{Rf}) to the actual positions of the
phonon levels "by hand", e.g., by applying the procedure
\begin{equation}
R \to R - \sum_{ \{ q,S_q \} }^{group}  \prod_q (-1)^{S_q}
\mid r(0,S_q,{2\pi \over LK\mid q \mid }) \mid ^2
\left(  { 1 \over 1- KM( \{q, S_q \} ) } -
{ E_1 \over E_1- E_{ph}( \{q, S_q \} ) }  \right) \;,
\label{Re}
\end{equation}
where the sum is over any group of levels characterized by $\{q, S_q \} $
with the matrix elements given by (\ref{ME})
(more precisely  the ratio has to be calculated from the secular equation for
the same group of levels). This looks like a reasonable approach
to the single-impurity case.
Still, in the disordered case we have further uncertainty
in the values of the matrix elements because of the integrals
$\int dx V_L(x) \exp \{ \pm i\gamma (x) \pm i2\pi mx/L \}$, which
depend on the particular  disorder realization. Obviously, a single
value of $\int dx V_L(x) \exp \{ i\gamma (x) \}$ taken from
the amplitude $\Delta_{AB}$ is not sufficient to fix all these integrals
for different phonon modes. In the general case we have to treat
these integrals as random values, which  for the lowest modes
are of order $\langle V_L \rangle $ in magnitude.
The avoided level crossing in Fig.2 tells unambiguously
that all the symmetries are  violated, and SC is coupled
to the left- and right-moving phonons with different matrix elements.
On the other hand we see clearly in Fig.8 that the
splitting $\Delta_1$ has extended resonance structure and the ratio $R$
is typically much larger than unity in qualitative agreement with the
theoretical prediction (\ref{Rf}).

Finally, we touch upon a problem of an extrapolation of
the finite-size exact-diagonalization results to the larger system sizes.
In 1D this problem is especially important since the finite-size cross-over
regions in the vicinities of the critical points have only logarithmical
macroscopic smallness. (For a typical example see the spectrum in Fig.\ 3
demonstrating quite a smooth behavior  at $V=2$.) Obviously, the extrapolation
can be performed with the RG equations. As the parameter $K$ naturally follows
from the spectrum, the only problem is to obtain $y$. This can be done by
fitting the behavior of $K$ as a function of $N_a$ with RG equations. The
details and results of such an approach see in \cite{KPS_2}.

\section{Possible realizations of 1D supercurrent states }
\label{sec:5}

In this section we discuss in some detail different systems and experimental
setups which allow to study SC states in a ring geometry.
The most obvious choice is a real 1D electron wire of mesoscopic size.
Recent experimental techniques \cite{exper} produce rings with
$L \sim 10^4\:A^o$  and only few transverse levels (in
ideal case one) below the chemical potential $\mu$. The experiments can be
done on ensembles as well as on single rings \cite{SRE}. In a static
experiment one may study  SC  by measuring the so-called
persistent current, which is defined as
\begin{equation}
j = -{1 \over 2\pi } \partial F / \partial \varphi \;\;\;\; \;\;\;(\varphi =
\phi / \phi_0 )\;,
\label{j}
\end{equation}
where $F$ is the free energy. At zero temperature $j=2v_I/L \dot
(I-\varphi )$,
where $I$ minimizes the potential energy $E_I(\varphi )$. In ideal system
$j(\varphi )$ is a saw-like function with abrupt change
of the current from $-v_I/L$ to $v_I/L$ at $\varphi =m+1/2$, when the two
energy levels $I=0$ and $I=1$ are degenerate.
The singularity is changed to
crossover when we incorporate the finite energy splitting $\Delta_{AB}$ into
the problem. A trivial solution of the two-state Hamiltonian gives now
\begin{equation}
j(\varphi ) = {2v_I\over L} \delta \varphi \left( { 2\pi v_I/L
\over \sqrt{(4\pi v_I\delta \varphi  /L )^2 +
\Delta_{AB}^2} } -1 \right) \;; \;\;\;
\delta \varphi = \varphi -1/2 \;.
\label{jd}
\end{equation}
We see that the crossover region is very  narrow
$\mid \delta \varphi \mid \sim L\Delta_{AB} /2v_I $ in the superfluid phase.
Clearly, such a ring may work as a very sensitive magnetometer capable
of measuring the flux with the accuracy $\sim \phi_0 (v_I /L\Delta_{AB} ) \ll
\phi_0$.
(An interesting experimental setup was proposed recently in \cite{Fazio},
where the ring conductance in the flux is probed by two weak tunnel junctions
at a distance $L/2$. In this case the Josephson critical current at zero
voltage changes dramatically at $\varphi =1/2$.)

The static picture just described is valid provided the time of experiment
is much longer than the relaxation time $\tau $ (see Eq.(\ref{tau})), or if
the flux was switched on adiabatically
$d \ln \varphi /dt \ll \Delta_{AB}^{-1}$. For large enough systems the both
requirements might be violated.  By inserting nonadiabatically the flux
$\varphi \approx m$ (same arguments go through for $\varphi=1/2$-integer)
we prepare the SC state $\mid I=m \rangle$. Then, according to the standard
two-state dynamics, the current starts oscillating with the frequency
$[(8\pi nv_I/L)^2 +\Delta_I^2]^{1/2}$. In a more sophisticated experiment
one may study the echo signal from the ring. Actually, the echo or
the burned hole experiment on the ensemble of rings with different sizes
seems to be the most appropriate one, because the wide spread in $\Delta_I$
and $\varphi $ for individual rings makes it difficult to see the resonant
absorption of the electro-magnetic field.
We are not going into details here  because this case is
identical (with clear modifications for the distribution function)
 to that of the two-level systems in amorphous materials (see e.g. the review
article \cite{Hunk}).

The other system described as a 1D Luttinger Liquid at very low
temperatures is in fact a 2D electron gas in the FQHE state. In the FQHE state
all bulk excitations are frozen at temperatures much less than the
energy gap in the collective spectrum of interacting electrons, and the only
low-energy degrees of freedom left in the incompressible electron liquid are
the edge excitations which form 1D chiral Luttinger Liquid
\cite{Wen,Been,MacD,Stone}. Here we will discuss the sample with the
Carbino disk geometry where the right- and left-moving fields are described
by the inner and outer edge currents correspondingly, and the relaxation
process is via quasiparticle tunneling between the edges through the bulk at
the point of constriction (or many constrictions if we are interested in the
disordered case). The crucial parameter $K^{-1}$ in this case equals to the
filling factor of the FQHE state $\nu $ \cite{Wen}, and for $\nu = 1/(2k+1)$
(when there is only one channel on each edge) we are in the insulator
phase, that is in the {\it macroscopic} sample $L \to \infty$ the constriction
is equivalent to cutting the ring. One may argue that our results do not
apply to this case because we concentrated on the superfluid phase only.
However in all our calculations done in sections \ref{sec:2}, \ref{sec:3} the
only assumption used was that the dimensionless vertex $w$ is small. In the
superfluid phase this was true even for a large microscopic "bare" vertex.
Suppose now that we have a {\it finite-size} ring with
extremely small "bare" value of $w$, so that even for large $K$
the scaling toward larger values of $w$ still leaves $w(L)$ small, and we
actually never enter the true insulating phase. In this case all our
results remain intact and directly apply to the experiment. Moreover,
large values of $K$ might be an advantage because we need a ring with the
distance between the edges $d$ much large than the magnetic length
$l_H$, in order to be able to speak about well-defined edge excitations
separated by inert bulk. The tunneling rate $w$
goes to zero exponentially with the distance $d$, and for not too narrow
disks will be unobservable if not for the increase of $w$ under
the scaling when $K$ is large.

The other advantage of the Hall system is that it is not sensitive to small
deviations of the magnetic field from special values,  and the
width of the disk $d$ does not constitute any problem when an external
magnetic field is varied to change the flux through the inner edge.
Obviously, the above discussion of possible experimental setups   for
studying the dynamics of supercurrent states is in order here. One can
also make use of the spatial separation between the left- and right-moving
currents (which is not the case in really 1D wires). In the Hall system
current relaxation is connected with the charge transfer between the edges,
which allows to follow, e.g., current oscillations, by monitoring the charging
effects in the vicinity of the sample.

In all examples discussed so far the parameter $K$ was larger than unity
(in 1D conducting wires due to Coulomb repulsion between the electrons we
expect $K >1$).
Consider now a 3D superconducting ring having the two spatial dimensions
$d_1$ and $d_2$ (defining the sample cross-section)
much smaller than the ring circumference $L$. The question we address here
is the size dependence of the supercurrent relaxation/oscillation
in such a ring. The niave answer is: since the relaxation is via vortex
line tunneling across the constriction (or vortex ring nucleation inside the
ring with subsequent expansion to the sample boundaries), then the
process will depend on $d_1$ and $d_2$, but not on $L$ because it
happens locally right at the point of deliberately introduced constriction
and none of the vortex sizes exceeds the cross-section diameter.
As we demonstrate below, this answer is absolutely wrong, and the relaxation
process depends explicitly on the ring length $L$.

At temperature
$T \ll T_c$ (we assume that the superconducting energy gap is nonzero and
of order $T_c$ all over the Fermi surface) we describe the superconductor
by effective action  for the phase of the order parameter
which has the same form as Eq.(\ref{S1})), except that now the integration over
$dx$ ought to be understood as $dx_1\: dx_2\:dx_3$. Let us go to
even lower temperatures $T \ll 1/md^2$ (we assume that the cross-section
diameter $d$ is larger than the coherence length $\xi$), when the transverse
excitation modes are frozen, and the effective action can be further
integrated over $dx_1$ and $dx_2$. What is left is the
familiar 1D effective action with $\sqrt{\Lambda_s /\kappa}  \sim v_F$ and
$\Lambda_s,\; \kappa \sim (d/a)^2$. Clearly, in this case we deal with
extremely small values of $K$
\begin{equation}
K \sim (a/d)^2 \ll 1,
\label{Kd}
\end{equation}
and the effective 1D system is indeed in the superfluid phase.
Thus we find that the rate of current oscillations will be
proportional to
\begin{equation}
\ln \Delta_{AB}/E_{AB} \sim  -(d/a)^2 \ln L/d
\label{rd}
\end{equation}
according to our general arguments which lead to (\ref{Eab}) (we
used the fact that $1/K \gg 1$).

Till now we completely ignored the effects of Coulomb interaction between
the particles, so the result (\ref{rd}) rather applies to the neutral
superfluid like He-II in a narrow long tube. One might also think
about the experiment where the superconducting ring is placed on top of
the insulator-metal plate, which will screen long-range Coulomb forces.
However if the interaction between the charges remains unscreened, the
result (\ref{rd}) has to be essentially modified.
The spectrum of collective modes in a 3D superconductor in the limit
$q \to 0$ is given by \cite{Anders} (see also \cite{Pop})
\begin{equation}
w_q^2 \sim v_F^2 q^2 \Pi(0) V_C(q)\;;
\label{wc}
\end{equation}
and for $V_C(q) \sim e^2/q^2 $ has a plasmon gap
$w_p^2=ne^2/m \sim v_F^2 \kappa_p^2$, where $\Pi(0)$ is the static
polarizability, and $\kappa_p^{-1}$ is the normal-metal screening radius.

Consider now the limit $L^{-1} \le q \ll d^{-1}$, which corresponds to our
quasi-1D geometry as discussed above. In this momentum range we almost
recover back the sound-like energy spectrum for phase fluctuations
\begin{equation}
w_q^2 \sim v_F^2 q^2 (\kappa_p d)^2 \ln qd \;.
\label{w1d}
\end{equation}
Weak logarithmic dependence of the sound velocity and parameter $K$ which
is now given as
\begin{equation}
K^{-1} \sim {d \over \kappa_p a^2}\:{1 \over \ln ^{1/2} (qd) }
\label{KCo}
\end{equation}
can not change the theory in any essential way, except that now the
integration of the RG equation $\int_{1/L}^{1/d} K^{-1} dq/q $ gives
$\ln ^{1/2} L/d $ instead of conventional logarithmic dependence. Thus
we find for the charged superconductor another law
\begin{equation}
\ln \Delta_{AB}/E_{AB} \sim  -{d \over \kappa_p a^2 } \ln ^{1/2} L/d \;.
\label{law2}
\end{equation}

Of course, to observe these dependences experimentally the ratio $d/a$
must be not too large. Bearing in mind that we assumed $d >\xi$, we
suggest superconductors with short coherence length (e.g., HTSC), or else
$^4$He-II, although we are not aware of any experimental technique
for preparing such narrow ($d \sim 20\:  \AA$) long tubes for the helium
experiment.

In this paper we mostly discussed the properties of SC in a sample
with the ring topology threaded by the flux. We realize, however,
that this situation can be modeled in a 1D wire connected to the
constant-current source through the Josephson contacts. In a way, here
one deals with a reverse problem of finding the phase difference
between the contacts as a function of current. This experimental setup
might be of particular interest for studing the inelastic relaxation
of SC due to their coupling  to the substrate at large values of the
current $j > j_c$, where $j_c$ is the maximum value of the persistent
current in Eq.(\ref{jd}). At mesoscopic values of the current $j \sim 1/L$,
the phase slippage process in a 1D system is possible only if assisted
by the energy dissipation to the substrate because we are dealing
with the lowest energy levels, and 1D phonons form a discrete spectrum
which is not in resonance with SC in the general case.

\section{acknowledgments}
We would like to thank Yu.Kagan and I. Affleck for the discussions and
valuable comments.

This work was supported by International Science Foundation (Grant No. MAA300)
(NVP and BVS) and by Russian Foundation for Basic Research
(Grants No. 94-02-05755a (VAK and AIP) and No. 95-02-06191a (NVP and BVS)),
and partially by Grants No. INTAS-93-2834 [of the European Community] (BVS)
and No. NWO-07-30-002
[of the Dutch Organization for Scientific Research] (NVP and BVS).

\figure{ \noindent {\bf Fig.1}.
Spectrum evolution of a disordered system with increasing disorder.
$N_a=11$, number of particles $N_b=7$, $U=2.5$. Here $(2,0)$ and $(0,2)$ are
the states with two identical minimal-momentum phonons; $(1,1)$ is the state
with two minimal-momentum phonons, moving in the opposite directions.
}

\figure{ \noindent {\bf Fig.2}.
Spectrum evolution of a disordered system with increasing interaction.
$N_a=11$, $N_b=7$, $W=0.5$. The notation is the same as in Fig.\ 1.
}

\figure{ \noindent {\bf Fig.3}.
Spectrum of the commensurate hard-core model as a function of the
nearest-neighbor interaction; $N_a=22$, $N_b=N_a/2$.
}

\figure{ \noindent {\bf Fig.4}.
Scaling of relative AB-splitting with Na in the case of single impurity
($\epsilon_1 =0.5$); $N_a = 8,10,12,14,16,18$; $N_b=N_a/2$.
}

\figure{ \noindent {\bf Fig.5}.
Scaling of relative AB-splitting with Na in the disordered case ($W=0.5$);
$N_a = 8,10,12,14,16,18$; $N_b=N_a/2$. Each point represents an average
over 100 realizations.
}

\figure{ \noindent {\bf Fig.6}.
Scaling of the relative splitting of the first SC with Na in the
commensurate case; $Na = 10,12,14,16,18,20,22$; $N_b=N_a/2$.
}

\figure{ \noindent {\bf Fig.7}.
Scaling of the relative splitting of the first SC with Na in the
commensurate case at the critical point.
$N_a = 10,12,14,16,18,20,22$, $N_b=N_a/2$.
}

\figure{ \noindent {\bf Fig.8}.
The ratio $R$ between the first SC energy splitting $\Delta_1$ and the
square of the lowest splitting  $\Delta_{AB}$;
$N_a=11$, $N_b=7$, and $W=0.25$.
}


\begin{thebibliography}{99}
\bibitem{we}
  The classification of levels adopted in \cite{Aff_et_al} is based on the
extra symmetry characteristic of the
isotropic Heisenberg Hamiltonian and needs,
generally speaking, some translation into the superfluid notation to make it
possible to compare the results with the general theory of 1D
superfluid-insulator transition. For one thing, the upper state of the first
SC doublet is referred to in \cite{Aff_et_al} as the lowest singlet
excitation and the lower state of the SC doublet corresponds to the lowest
triplet excitation of \cite{Aff_et_al} (the one with zero $z$-projection of
the total spin).
\end{thebibliography}
\end{document}